\documentclass[%
 reprint,
superscriptaddress,
 amsmath,amssymb,
 aps,
]{revtex4-2}
\usepackage{textcomp}

\usepackage{graphicx}
\usepackage{dcolumn}
\usepackage{bm}
\usepackage{gensymb} 
\usepackage[utf8]{inputenc}
\usepackage{ragged2e}
\usepackage{subfig}
\usepackage{xcolor}
\usepackage{verbatim}
\usepackage[normalem]{ulem}

\usepackage{xr-hyper}
\usepackage{hyperref}
\usepackage{MnSymbol}
\usepackage{float}
\usepackage{amsmath}
\usepackage{xspace}

\newcommand{\rfr}{\ensuremath{_{\mbox{\small ref}}}\xspace}
\newcommand{\SB}[1]{\ensuremath{\mbox{SB}{#1}}\xspace}
\newcommand{\HA}[1]{\ensuremath{\mbox{H}{#1}}\xspace}
\newcommand{\delay}{\ensuremath{\tau_{\mbox{\small tran}}}\xspace} 

\definecolor{Green_Dominique}{rgb}{0, 0.5, 0}

  \hypersetup{
  colorlinks = true,
  linkcolor = purple,
  citecolor = red,
urlcolor = blue!75!black
  }

\definecolor{rtcol}{rgb}{0.,0.,1.}

\definecolor{jccol}{rgb}{1.,0.13,0.32}

\definecolor{aacol}{rgb}{0.,0.8,0.2}

\definecolor{pscol}{rgb}{1.,0.13,0.92}

\begin{document}

\hyphenation{pro-ba-bi-li-ties}
\hyphenation{mo-du-la-tion}
\hyphenation{stu-dies}
\hyphenation{pho-to-io-ni-za-tion}
\hyphenation{using}
\hyphenation{qual-i-ta-tive-ly}
\hyphenation{analy-sis}
\hyphenation{theo-re-ti-cal-ly}

\preprint{APS/123-QED}

\title{
Anisotropic dynamics of two-photon ionization: \\An attosecond movie of photoemission
}

\newcommand{\LIDYL}{Universit\'e Paris-Saclay, CEA, CNRS, LIDYL, 91191 Gif-sur-Yvette, France}
\newcommand{\LCPMR}{Sorbonne Universit\'e, CNRS, Laboratoire de Chimie Physique-Mati\`ere et Rayonnement, 75005 Paris, France}
\newcommand{\ILM}{Universit\'e de Lyon, Universit\'e Claude Bernard Lyon 1, CNRS, Institut Lumi\`ere Mati\`ere, 69622 Villeurbanne, France}
\newcommand{\ISMO}{Universit\'e Paris-Saclay, CNRS, Institut des Sciences Mol\'eculaires d'Orsay, 91405 Orsay, France}

\author{A.~Autuori*}
\author{D.~Platzer*}
\author{M.~Lejman}
\author{G.~Gallician}
\author{L.~Ma\"{e}der}
\author{A. Covolo}
\author{L.~Bosse}
\author{M.~Dalui}
\author{D.~Bresteau}
\affiliation{\LIDYL}
\author{J.-F.~Hergott}
\author{O.~Tcherbakoff}
\affiliation{\LIDYL}
\author{H. J. B.~Marroux}
\affiliation{\LIDYL}
\author{V. Loriot}
\affiliation{\ILM}
\author{F. L\'epine }
\affiliation{\ILM}
\author{L.~Poisson}
\affiliation{\LIDYL}
\affiliation{\ISMO}
\author{R.~Ta\"{i}eb}
\author{J.~Caillat}
\affiliation{\LCPMR}
\author{P.~Sali\`eres}
\affiliation{\LIDYL}

\date{\today}
\begin{abstract}

 Imaging in real time the complete dynamics of a process as fundamental as photoemission has long been out of reach due to the difficulty of combining attosecond temporal resolution with fine spectral and angular resolutions. Here, we  achieve full decoding of the intricate angle-dependent dynamics of a photoemission process in helium, spectrally and anisotropically structured by two-photon transitions through intermediate bound states. Using spectrally- and angularly-resolved attosecond electron interferometry, we characterize the complex-valued transition probability amplitude towards the photoelectron quantum state. This allows reconstructing in space, time and energy the complete formation of the photoionized wavepacket. 

\end{abstract}

\maketitle

Photoemission has played a key role in the development of quantum physics theory. Beside its fundamental aspects, it has become a widespread technique for the investigation of matter, from chemical analysis in condensed matter to electronic structure determination in the gas phase.
The development of bright sources of extreme ultraviolet (XUV) radiation such as synchrotrons has allowed the measurement of angular distributions of photoelectrons with high precision both in the laboratory and molecular frames. Pioneering complete experiments \cite{Cherepkov_2000,LebechJPC2003} gave access to the relative weights and phases of the partial waves contributing to the wavefunction of the liberated electron. Such measurements do not give access to photoemission in the time domain because they are performed {\it independently} for each final electron energy and thus lack the relative phase between the spectral components. 

The advent of broadband coherent XUV sources based on high-order harmonic generation (HHG) from intense infra-red (IR) pulses \cite{McPherson_JOSAB_1987,Ferray_JPB_1988} has triggered the development of unique interferometric schemes to measure spectral phases, leading to unprecedented insight into the temporal dynamics of elementary processes in a broad range of chemical species with attosecond resolution (1 as $=10^{-18}$ s). Both the streaking \cite{Hentschel_Nature_2001} and the Reconstruction of Attosecond Beating By Interference of two-photon Transitions (RABBIT) technique \cite{PaulScience2001,Mairesse_2003} were applied to revisit photoemission in the time domain, through  the first measurements of attosecond photoemission delays in atoms \cite{SchultzeScience2010,KlunderPRL2011}, molecules \cite{HaesslerPRA2009,Huppert_2016}, nanoparticles \cite{Seiffert_2017}, liquids \cite{Jordan_2020} and solids \cite{CavalieriNature2007}. Access to higher levels of details was gradually achieved using tunable sources~\cite{SwobodaPRL2010,Schoun_PRL_2014,Drescher_ArXiv_2021} or by increasing the spectral resolution at detection with  the Rainbow RABBIT technique \cite{GrusonScience2016,Isinger_Science_2017}. 
However, all the aforementioned studies relied on measurements averaged over the direction of photoemission. The reconstructed dynamics thus remained {\em incomplete}. 
While attosecond photoelectron interferometry has been investigated in momentum space in earlier experiments \cite{AseyevPRL2003,RemetterNatPhys2006,Mauritsson_PRL_2010}, it is only recently that orientation-resolved 
spectral phase measurements could  be performed, using cold target recoil ion momentum spectroscopy (COLTRIMS) \cite{HeuserPRA2016,CirelliNC2018,JosephJPhysB2020,FuchsOptica2020}, or velocity-map imaging spectroscopy (VMIS) \cite{BeaulieuScience2017,Villeneuve_Science_2017,BustoPRL2019}. 

In the present study, we combine attosecond spectral interferometry with momentum spectroscopy to record the modulus and phase variations of the photoelectron quantum state with high spectral resolution and angular sensitivity. The  potential of this complete quantum phase spectroscopy is demonstrated in the test case of two-photon XUV+IR photoionization of helium through the intermediate resonant states $1s3p$ and $1s4p$. The resulting structured photoelectron wavepacket is fully characterized by measuring quasi-continuously the  spectral and spatial variations of the final quantum state over a 0.8-eV spectral range. This allows reconstructing the attosecond photoemission dynamics  strongly impacted by the sudden phase jumps of up to $\pi$ rad measured {\em in both dimensions}.

\begin{figure}[h]
\includegraphics[width=1.0\linewidth]{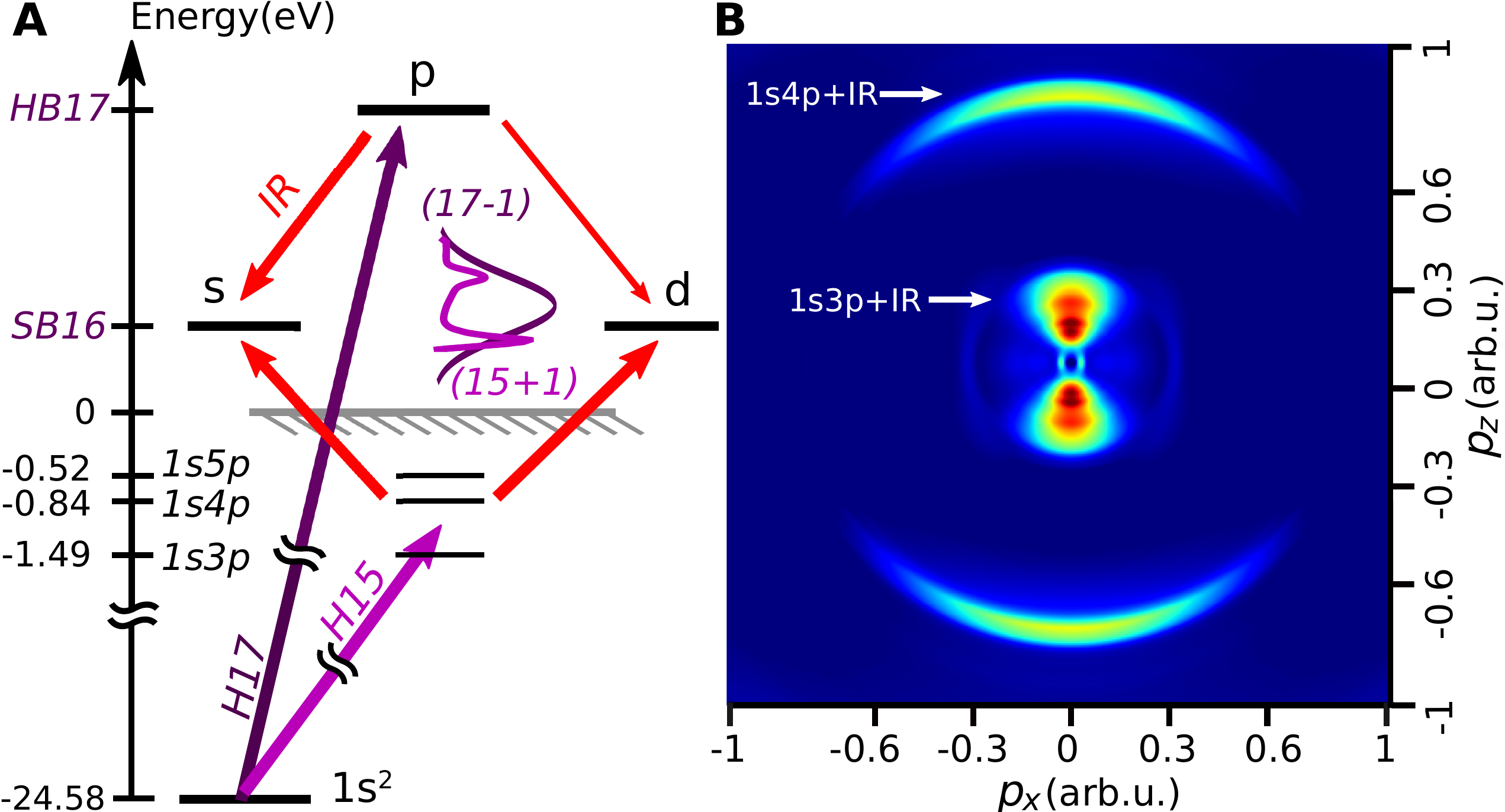}
\caption{\textbf{Principle of spectrally- and angularly-resolved photoelectron interferometry. (A)} 
Helium is ionized with a comb of coherent odd harmonics of a Titanium:Sapphire laser combined with a weak fraction of the fundamental laser. The two-photon electron wavepacket created in the  continuum by absorption of harmonic 15 (\HA{15}) and an infrared (IR) photon is highly structured due to the intermediate $1s3p$ and $1s4p$ resonant states. This wavepacket, (15+1), is a coherent superposition of the $s$ and $d$ partial waves that are populated by the two-photon transition. A reference wavepacket,  (17-1), is created at the same energy by absorption of \HA{17} and stimulated emission of an IR photon.
\textbf{(B)} Momentum map of the delay-integrated \SB{16} from an Abel-inverted VMIS measurement. It exhibits two main spectral components centered around $1s3p$ + IR and $1s4p$ + IR.}
\label{fig:He_raw_with_levels_eng_sym_image}
\end{figure}

The concept of the technique is illustrated in Fig. \ref{fig:He_raw_with_levels_eng_sym_image}
\begin{figure*}[]
\hspace*{-1.0cm}
\includegraphics[width=1.1\linewidth]{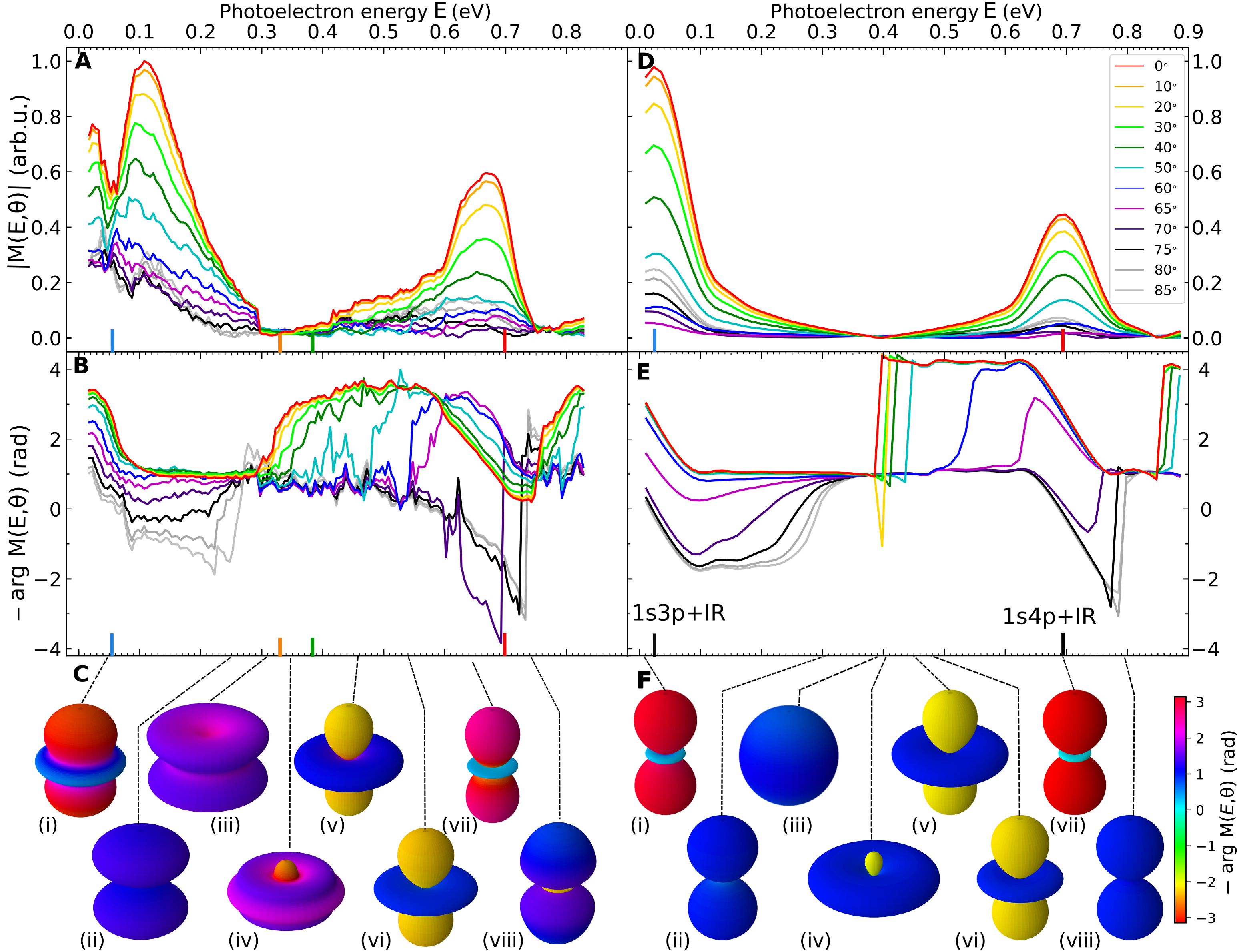}
\caption{\textbf{Probability  amplitude associated with the investigated (15+1) two-photon transition.} Spectral and angular variations of the modulus and phase of $M(E,\theta)$ resulting from a Rainbow RABBIT analysis of the measured (resp. \textbf{A} and \textbf{B}) and simulated (resp. \textbf{D} and \textbf{E}) spectrogram $I_{16}(\tau; E, \theta)$. The phases are shown with a negative sign for readability. Polar representation of $M(E,\theta)$ at specific energies labeled from (i) to (viii), where the $\theta$-dependent radius and color correspond respectively to the modulus and phase from experimental measurements (\textbf{C}) and from TDSE simulations (\textbf{F}).} 
\label{fig:RABBIT_0_90}
\end{figure*}
We aim at fully characterizing the resonant (15+1) two-photon transition in He using a reference (17-1) transition. Their interference determines the photoelectron angular and spectral distribution in the sideband peak (\SB{16}) that is measured by VMIS (see Fig. \ref{fig:He_raw_with_levels_eng_sym_image}B). The \SB{16} intensity $I_{16}$ thus depends on 3 parameters: i) the XUV-IR delay $\tau$; ii) the photoelectron energy $E$; and iii) the angle  $\theta$ between the electron emission and the shared XUV-IR polarization axis $\textbf{z}$, in the generic form:
\begin{equation}
I_{16}(\tau; E, \theta) 
  =  A_{16}(E, \theta) + B_{16}(E,\theta) \cos [2\omega_0 \tau - C_{16}(E, \theta)],
 \label{SB_VS_tau_and_theta}
 \end{equation} 
where $\omega_0$ is the laser frequency.  
Our study focuses on the amplitude $B_{16}$ and phase $C_{16}$, out of which the complex anisotropic photoemission dynamics is decoded.
Both quantities are accessed through a Fourier transform of the 3D spectrogram $I_{16}(\tau; E, \theta)$ with respect to $\tau$ at each sampled energy $E$.
They are then calibrated in order to extract the intrinsic two-photon transition amplitude $M(E,\theta)$ associated with the probed $(15+1)$ path independently of the characteristics of the XUV exciting pulses. 

The modulus and phase of $M(E,\theta)$  obtained from the experimental data are shown in Fig.~\ref{fig:RABBIT_0_90}A-B. They display many structures related to the build-up of the wavepacket through intermediate resonances. At $E\approx E_{1snp}+\hbar\omega_0$, the modulus is enhanced and a $\approx\pi$ rad smooth spectral phase drop occurs, reminiscent of the observations of Ref.~\cite{SwobodaPRL2010}. In contrast, between resonances, the signal significantly drops and a  sharp $\approx\pi$ rad spectral phase jump occurs. A most remarkable feature is the strong angular dependence of both modulus and phase (jumps up to $\pi$ rad)  over the covered spectral range. This is highlighted  by fast changes in shape and phase of the polar representation of $M(E,\theta)$, as shown in Fig.~\ref{fig:RABBIT_0_90}C for few selected representative energies.

We performed numerical simulations based on the Time Dependent Schr\"odinger Equation (TDSE) with a model potential for helium and pulse characteristics corresponding to the experimental ones. The results, shown in Fig.~\ref{fig:RABBIT_0_90}D-E, are in very good agreement with the experimental ones,
the latter showing slightly attenuated phase jumps that are down-shifted by $\sim 80$~meV around $E=0.32$~eV, and a stronger angular dependence around the $1s4p$ resonance. 

The observed rich spectral and angular features are the signature of a structured anisotropic photoemission dynamics, fully encoded in the differential transition probability amplitudes $M(E,\theta)$. They can be understood through a partial wave expansion of the latter, 
\begin{eqnarray}\label{eqn:PWexp}
M(E,\theta)&=&\mathcal{M}_0(E)Y_{00}(\theta)+\mathcal{M}_2(E)Y_{20}(\theta),
\end{eqnarray}
where $Y_{\ell0}$ are the spherical harmonics, and $M_{\ell}(E)$, the matrix elements associated with each final angular momentum $\ell$.

\begin{figure*}
\hspace*{-0.5cm}
\includegraphics[width=0.8\linewidth]{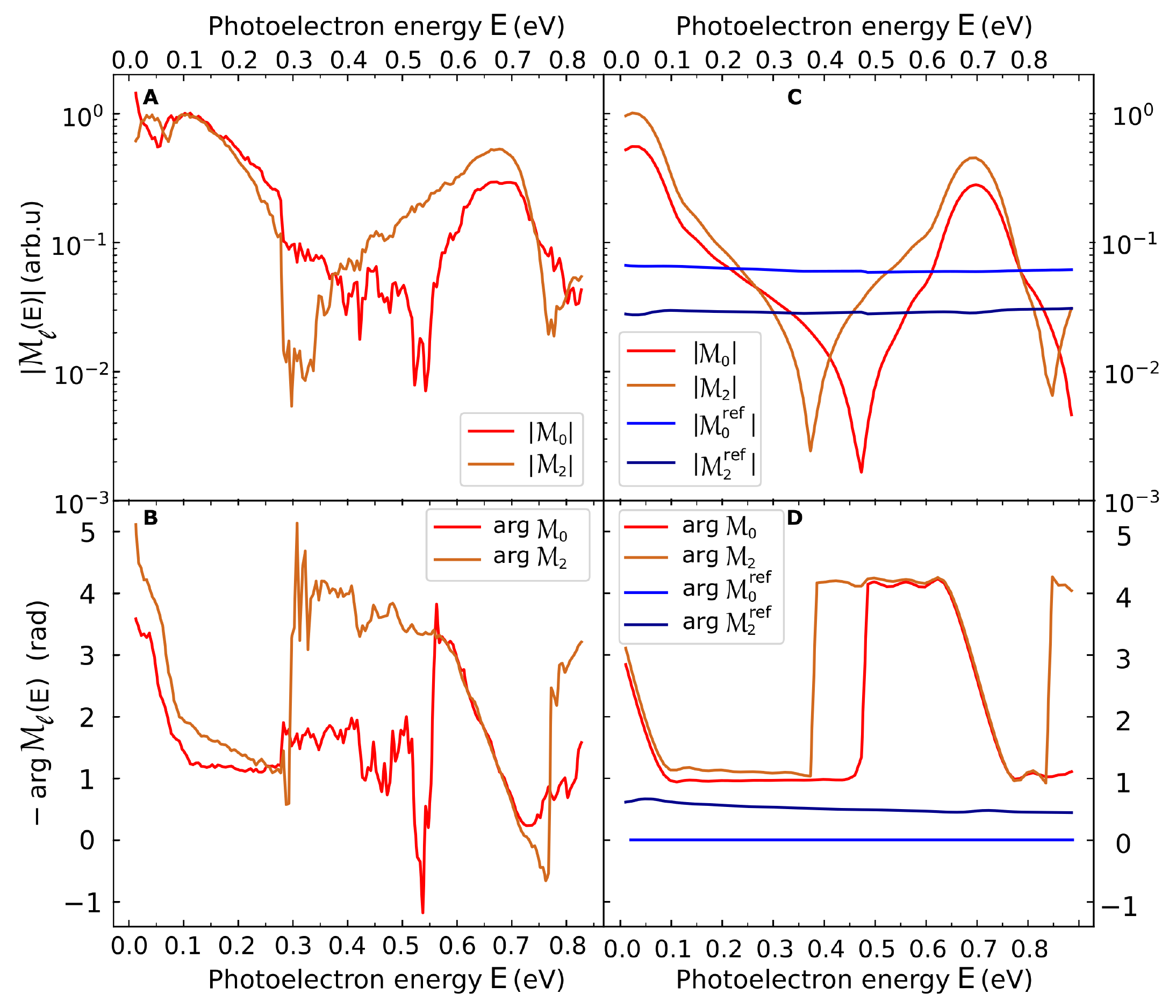}
 \caption{\textbf{Partial wave decomposition of the investigated two-photon transition}. Modulus and phase of the partial wave transition amplitudes $\mathcal{M}_{\ell}(E)$ ($\ell=0,2$) obtained from experiments (resp. \textbf{A} and \textbf{B}) and from simulations (resp. \textbf{C} and \textbf{D}). The simulated partial wave amplitudes for the reference transition are also shown in \textbf{C-D}, with $\arg \mathcal{M}^{\mbox{\small ref}}_0(E)$ taken as phase origin. }
\label{fig:Figure_th_sd}
\end{figure*}

For monochromatic fields, in our spectral range where the XUV photon frequency $\Omega$ is nearly resonant with the $1snp$ states, the partial amplitudes can be approximated by: 
\begin{equation}
{\cal M}_{\ell} (E) \approx  \lim\limits_{\epsilon \to 0^{+}} \sum\limits_{n=3}^{5}\frac{\langle 1s E\ell|{z}| 1snp \rangle \langle 1snp|{z}| 1s^2 \rangle} {E_{1s^2} - E_{1snp} + \hbar\Omega + i\epsilon  },
 \label{Matrix_element_at_res}
 \end{equation} 
where $| 1s^2 \rangle$ is the initial ground state of energy $E_{1s^2}$, $\vert 1s E\ell\rangle$ is the considered final partial wave with photoelectron energy $E=E_{1s^2} + \hbar\Omega+\hbar\omega_0$. In our experiments and simulations, the finite duration/bandwidth of the XUV and IR fields results in a smoothing of the outcomes from Eq.~\ref{Matrix_element_at_res} \cite{JimenezPRA2016,VacherJO2017}, which accounts for the spectral variations of $M_\ell(E)$ shown in Fig.~\ref{fig:Figure_th_sd}C-D.


In contrast to the structureless (17-1) reference path, dominated by the isotropic $s$ wave, the studied transition matrix element for each of the $\ell$ channels undergoes a series of $\pi$-rad phase jumps accompanied by a strong modulus increase, around 0.06 and 0.71~eV photoelectron energies. They are reminiscent of a zero crossing of the denominator in Eq.~\ref{Matrix_element_at_res}, wherever $\Omega$ is resonant with one of the $1snp$ ($n=3,4$) levels.
Furthermore, between consecutive resonances, two  contributions with opposite signs dominate the sum, and cancel each other at a given energy, depending on  the relative values of the $\ell$-dependent numerators. 
This results in a local minimum in the transition amplitude and a very sharp $\pi$ phase jump at 0.47~(0.37)~eV for the $s$ ($d$) channel. These $\ell$-specific cancellations are highly sensitive to the resonance positions and relative strength. They are  measured at 0.54~(0.30)~eV for the $s$ ($d$) channel in the experimental decomposition in Fig.~\ref{fig:Figure_th_sd}A-B, that otherwise displays a very good agreement with the theoretical prediction.

The relative imprint of these $\ell$ contributions over the whole energy range is clearly visible  in the polar representation of the final quantum state in  Fig.~\ref{fig:RABBIT_0_90}C-F, which again shows a good correspondence between experiments and simulations, and slight shifts in energy. In the $1s3p$ resonance region, the dominant $d$ wave results in a minimum and change of sign at an angle $\theta_0\approx 60\degree$ \cite{BustoPRL2019,KheifetsPRA2021} (red to cyan in plot i), close to the node of  $Y_{20}(\theta)$. As the energy increases, the $d$ amplitude becomes less dominant, $\theta_0 \longrightarrow 90\degree$ and the sign change disappears (ii). The sharp $d$-wave cancellation results in a (almost) spherical $s$ state  (iii), experimentally distorted by a residual $\mathcal{M}_4$ contribution. Conversely, the $s$-wave cancellation leads to a typical $d$ state (vi). Between these energies, the $\pi$ relative phase of the s and d components produces a strong destructive interference at small angles, resulting in a doughnut shape (iv) evolving towards a $d$ shape (v). In the $1s4p$ resonance region (vii-viii), one recovers the initial shapes (i-ii). 

Our measurements  give direct access to the complete, angularly-resolved, dynamics of the two-photon transition leading to photoemission. A way of characterizing this dynamics is provided by the transition delay,
 a quantity specific to multiphoton processes defined as the {\em local} spectral derivative of the transition phase (see \cite{VacherJO2017} and references therein), for each emission angle $\theta$:
\begin{eqnarray}
    \delay(E,\theta)= \frac{\partial \arg M(E,\theta)}{\partial E}.
\label{phase_derivative}
\end{eqnarray}

\begin{figure}[]
\centering
\includegraphics[width=0.85\linewidth]{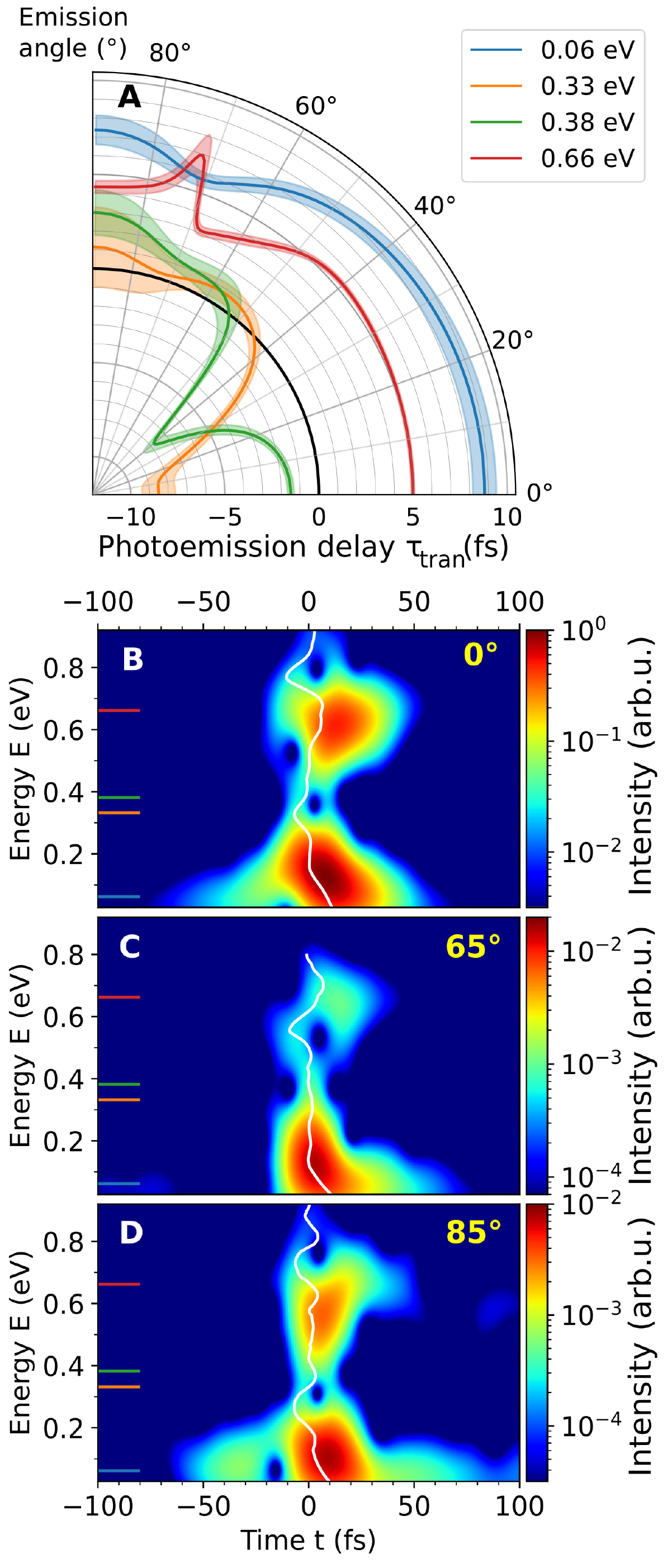}
\caption{\textbf{Photoemission dynamics retrieved from the experimental amplitudes $M(E,\theta)$. (A)} Angular dependence of the photoemission delay \delay determined by linear fit over 40 meV around four representative energies indicated by color ticks in Fig.~\ref{fig:RABBIT_0_90}, and corresponding uncertainty given by the fit standard deviation. \textbf{(B-D)} Spectro-temporal build-up of the  process revealed by a Gabor analysis of $M(E,\theta)$ at  three representative angles. The transition delay \delay as a function of $E$ is overlaid on each Gabor representation  (white lines).
}

\label{fig:Gabor_0_60_80_deg_delay}
\end{figure}

The results, shown in Fig. \ref{fig:Gabor_0_60_80_deg_delay}A reveal that, in the resonance regions, the delay is strongly positive and weakly anisotropic, while in the intermediate region, it is strongly negative and varies a lot with angle. 
This demonstrates that an angular variation of the {\em phase} should not be read as an angular variation of the {\em delay}, as a misinterpretation of the standard -- but here inappropriate -- atomic delay $\tau_{\mbox{\sc a}}(E,\theta) = [\arg M\rfr(E,\theta)-\arg M(E,\theta)]/2\omega_0$  would suggest.
The physical interpretation of this transition delay is straightforward  at each resonance: It represents an effective time during which the electron is transiently trapped in the intermediate bound state, before completing the transition~\cite{VacherJO2017}.
Its effective value is bounded by the experimental duration of the IR, which acts as a temporal gate on the process. In the intermediate region, {\em between} resonances, the interpretation is not so intuitive since here the dynamics are exclusively shaped by destructive quantum interferences.

To get a better insight on how the different spectral components interfere in the temporal build-up of the photoelectron wave-packet, we perform a Gabor analysis of the experimental $M(E,\theta)$, using a sliding 210-meV Gaussian gate as a balanced spectro-temporal resolution compromise.
The results for 3 illustrative angles are shown in Fig. \ref{fig:Gabor_0_60_80_deg_delay}B-D, overlaid with \delay now as a function of $E$.
As anticipated, the revealed dynamics  appear to be globally ``delayed'' at resonances (i.e. around 0.06 and 0.66 eV). Apart from this, they are strongly shaped by the presence of the spectral phase jumps distributed over the covered range. They result in destructive interferences in the temporal profiles, appearing as 'holes' in the Gabor transform. 
In particular, the 'holes' in the intermediate region between resonances shift towards higher energies when $\theta$ increases, as they follow the spectral phase jump (e.g., from 0.33~eV at $0\degree$ to 0.59~eV at $65\degree$). The spectro-temporal structuration of the dynamics therefore highly depends on the photoemission angle.

Our study thus establishes that it is now possible to follow the whole angular and temporal evolution of a structured wavepacket build-up during photoionization, making the dream of 'photoemission 3D movie' come true. By accessing simultaneously the spectral and angular phase variations, the method allows detailed studies in both dimensions, and in particular, gives direct access to the exact photoemission delays as the local derivative of the spectral phase, evidencing the shortcomings of the standard approximate expression. 
Most importantly, our straightforward analysis of the measured data to retrieve the complex transition amplitude is fully experimental (including the calibration procedure), and thus requires no theoretical input. This makes it applicable to a broad variety of atoms, (possibly laser-aligned) molecules and even solid state systems using hemispherical analyzers. The detailed information on the quantum photoemission processes will provide a stringent test for theories, in particular aiming at describing correlated ultrafast multi-electronic and vibronic dynamics, a general endeavor common to many fields.
Finally, the advanced characterization of Resonance-Enhanced Multi-Photon Ionization provided here opens new prospects for the widely-used REMPI technique \cite{dessent_CR_2000} in a broad range of applications and chemical systems.

\newpage
\bibliography{arXiv_biblio.bib}

\section*{Acknowledgments}
This research was supported by Agence Nationale de la Recherche, Grants No. ANR-20-CE30-0007-02-DECAP, No. ANR-15-CE30-0001-CIMBAAD, No. ANR-11-EQPX0005-ATTOLAB, No. ANR-10-LABX-0039-PALM; COST, Grant No. CA18222-AttoChem and Laserlab-Europe, Grant No. EU-H2020-871124.

AA and DP contributed equally to this work. AA, DP, ML, GG, LM, AC, LB, MD, DB, OT, JFH and LP performed the experiments. AA, DP, ML, AC and PS analyzed the data.  VL, FL and HJBM made independent measurements. RT and JC performed the simulations and theoretical analysis. AA, PS, JC, RT wrote the original draft. All authors contributed to the final version. PS, LP, FL, RT, JC obtained the funding that supported this work.

\end{document}